# Mobile Web Service Discovery in Peer to Peer Networks

Satish Narayana Srirama[1], Matthias Jarke[1,2], and Wolfgang Prinz[1,2]

[1] Information Systems Group, RWTH Aachen University
Ahornstr. 55, 52056 Aachen, Germany
[2] Fraunhofer FIT
Schloss Birlinghoven, 53754 Sankt Augustin, Germany
{srirama, jarke}@cs.rwth-aachen.de
wolfgang.prinz@fit.fraunhofer.de

**Abstract.** The advanced features of today's smart phones and hand held devices, like the increased memory and processing capabilities, allowed them to act even as information providers. Thus a smart phone hosting web services is not a fancy anymore. But the relevant discovery of these services provided by the smart phones has became quite complex, because of the volume of services possible with each Mobile Host providing some services. Centralized registries have severe drawbacks in such a scenario and alternate means of service discovery are to be addressed. P2P domain with it resource sharing capabilities comes quite handy and here in this paper we provide an alternate approach to UDDI registry for discovering mobile web services. The services are published into the P2P network as JXTA modules and the discovery issues of these module advertisements are addressed. The approach also provides alternate means of identifying the Mobile Host.

## 1  Introduction

The next generation devices like smart phones, PDAs and other communication gadgets are quickly filling up the cellular market today, creating endless possibilities for wireless communication. Demand for related software applications is also skyrocketing. Furthermore, recent developments in mobile communication technologies with 3G [1] and 4G [2] technologies have significantly increased the wireless data transmission rates achieved. All these developments together have made the web services usage a practical reality in cellular domain.

In the mobile web services domain, the resource constrained mobile phones are used as both web service clients and providers, still preserving the basic web services architecture in the wireless environments. While mobile web service clients are quite common [3], and many software tools [4], [5] are already existent in the market, easing their development and adoption, we have studied the scope of mobile web service provisioning in one of our previous projects. In this project, we have developed a Mobile Host [6], capable of providing basic web services from smart phones. Extensive performance analysis of the Mobile Host was conducted and many applications were designed and developed proving the feasibility of concept.

While the applications possible with mobile web services are quite welcoming, the huge number of web services possible, with each Mobile Host providing some services in the wireless network, makes the discovery of these services quite complex. Proper discovery mechanisms are required for successful adoption of mobile web services into commercial environments. The traditional centralized UDDI based registries [7] have many limitations in this aspect and might not be the perfect solution for the mobile web service discovery. The dynamic nature of the mobile nodes further enhances this problem.

In this paper we are proposing an alternative means of discovering mobile web services. The method uses the peer to peer (P2P) [8] network for advertising the web services and depends on network for discovering the services. We have developed the solution using the JXTA network and its features, and were able to publish mobile web services and discover them from the smart phones, with reasonable performance latencies. In this paper we will provide our approach, evaluation results and share our experiences. The rest of the paper is organised as follows.

Section 2 discusses the concepts of mobile web services, provisioning and the mobile web service discovery issues. Section 3 discusses the concept of mobile web services in P2P networks, addressing the convergence of mobile web services and P2P systems. Section 4 discusses the approach of publishing and discovery of mobile web services in JXTA network and provides our evaluation results. Section 5 concludes the paper and proposes future research directions.

## 2 Mobile Web Services

Service Oriented Architecture (SOA) describes a new component model which relates distributed components, called services, to each other by means of formally defined interfaces [9]. Thus SOA provides loose coupling of cleanly encapsulate services. Usually, SOA is implemented by means of web services which enable application-to-application communication over the Internet. SOA and Component-orientation are not new and SOA can also be realized with technologies like Common Object Request Broker Architecture (CORBA) [10] and DCOM (Distributed Component Object Model) [11]. But using web services for SOA provides certain advantages over other technologies. Web services are based on a set of still evolving, though well-defined W3C standards, that allow much more than, just defining interfaces.

Web services are self-contained, modular applications with their public interfaces defined and described using Web Services Description Language (WSDL) [12]. Web services provide access to software components through protocols like SOAP over different transmission protocols like HTTP, BEEP, UDP and etc [13], [14]. A service provider develops and deploys the service and publishes its description and access details (WSDL) with the UDDI registry. Any potential client, who queries the UDDI, gets the service description and accesses the service using SOAP. The quest for enabling these open XML web service interfaces and standardized protocols also on the radio link, with the latest developments in cellular domain, lead to a new domain of applications, mobile web services. The developments in cellular world are two folded; firstly there is a significant improvement in device capabilities like better memory and processing power and secondly with the latest developments in mobile communication technologies with 3G and 4G technologies, higher data transmission rates in the order of few mbs were achieved.

### 2.1 Mobile Web Service Provisioning

In the mobile web services domain, the resource constrained mobile devices are used as both web service clients and providers, still preserving the basic web services architecture in the wireless environments. While mobile web service clients are quite common these days, the research with providing web services from smart phones is still sparse. In our mobile web service provisioning project one such Mobile Host was developed proving the feasibility of concept.

Mobile Host is a light weight web service provider built for resource constrained devices like cellular phones. It has been developed as a web service handler built on top of a normal Web server. The SOAP based web service requests sent by HTTP tunneling are diverted and handled by the web service handler component. The Mobile Host was developed in PersonalJava on a SonyEricsson P800 smart phone [6]. The footprint of the fully functional prototype is only 130 KB. Open source kSOAP2 [15] was used for creating and handling the SOAP messages.

The detailed evaluation of this Mobile Host clearly showed that service delivery as well as service administration can be done with reasonable ergonomic quality by normal mobile phone users [16]. As the most important result, it turns out that the total web service processing time at the Mobile Host is only a small fraction of the total request-response time (<10%) and rest all being transmission delay. This makes the performance of the Mobile Host directly proportional to achievable higher data transmission rates. Similar implementations of Mobile Host were also possible with other Java variants like J2ME [17], for smart phones. We also have developed a J2ME based Mobile Host and its performance was observed to be not so significantly different from that of the PersonalJava version.

Mobile Host opens up a new set of applications and it finds its use in many domains like mobile community support, collaborative learning, social systems and etc. Primarily, the smart phone can act as a multi-user device without additional manual effort on part of the mobile carrier. Many applications were developed and demonstrated, for example in a distress call; the mobile terminal could provide a geographical description of its location (as pictures) along with location details. Another interesting application scenario involves the smooth co-ordination between journalists and their respective organizations [16]. Most recently the scope of the Mobile Host in m-learning (mobile learning) domain is also being studied with applications like podcasting, mobile blogging, expertise finders and etc. The Mobile Host in a cellular domain is of significant use in any scenario which requires polling that exchanges significant amount of data with a standard server, for example a mobile checking for the updates of RSS feeds provided by a server. The Mobile Host can eliminate polling process as the RSS feeds can now be directly sent to the Mobile Host, when the RSS feeds updated [18].

From the commercial viewpoint, there can be a reversal of payment structures in the cellular world. While traditionally the information-providing web service client has to pay to upload his or her work results to a stationary server (where then other clients have to pay again to access the information), in the Mobile Host scheme responsibility for payment can be shifted to the actual clients -- the users of the information/services provided by the Mobile Host. Thus Mobile Host renders possibility for small mobile operators to set up their own mobile web service businesses without resorting to stationary office structures [6].

While the applications possible with mobile web services are quite welcoming, the huge number of web services possible, with each Mobile Host providing some services in the wireless network, makes the discovery of these mobile services in a cellular domain, quite complex.

### 2.2 Discovery Issues

Generally, the WSDL document, that defines and describes a web service, consists of information specific to the location of the service (binding information) and the operations (methods) the service exposes. The web services are published by advertising these WSDL documents in a UDDI registry. The registry maintains a reference of the WSDL documents. Any potential web service client searches for the service in the public registry, gets the description of the service and tries to access the service using the information specified by the WSDL. Since the Mobile Host is implemented on the smart phone, mostly by using the basic web services architecture, the standard WSDL and UDDI registry can theoretically be used to describe and publish the services. Obtaining the binding information of the mobile web services can be tricky as it needs the IP address of the Mobile Host, where the services are deployed. Different means of accessing the services deployed on the Mobile Host are observed in [16].

But in a commercial environment with Mobile Hosts, and with each Mobile Host providing some services in the wireless network, the number of services expected to be published could be quite high. In such a situation, a centralized solution is not a best idea, as they can have bottlenecks and can make single points of failure. Besides, mobile networks are quite dynamic due to the node movement. Nodes can join or leave network at any time and can switch from one operator to another operator. This makes the binding information in the WSDL documents, inappropriate. Hence the services are to be republished every time the Mobile Host changes the network. This process leaves many stale advertisements in the registry. Keeping up to date information of the published mobile web services in centralized registries is really difficult.

Hence we are studying alternate means of discovering the web services deployed with Mobile Hosts. Here we will be addressing our proposed solution using P2P networks. Before explaining the discovery of mobile web services in P2P network, we discuss the issues with Mobile Host's entry into P2P network.

## 3   Mobile Web Services in P2P Networks

During our application analysis of Mobile Host, it was observed that most of the targeted collaborative applications, somehow converged to P2P applications and P2P offered a large scope for many applications with Mobile Host. P2P is a set of distributed computing model systems and applications used to perform a critical function in a decentralized manner. Peers are autonomous and in its pure form; each peer acts as both server and client. P2P takes advantage of resources of individual peers like storage space, processing power, content and achieves scalability, cost sharing and anonymity, and thereby enabling ad-hoc communication and collaboration. P2P systems have evolved across time and have wide range of applications and provide a good platform for many data and compute intensive applications [19], [20].

In order to adapt the Mobile Host to the P2P network, many of the current P2P technologies like Gnutella [19], Napster and Magi [21] are studied in detail. Most of these technologies are proprietary and are generally targeting specific applications. Only Project JXTA [22] offers a language agnostic and platform neutral system for P2P computing. JXTA technology is a set of open protocols that allow any connected device on the network ranging from cell phones and wireless PDAs to PCs and servers to communicate and collaborate in a P2P manner. JXTA enables these devices running on various platforms not only to share data with each other, but also to use functions of their respective peers. JXTA peers use XML as standard message format and create a virtual P2P network over these devices connected over different networks.

Combining JXTA and web services is not a completely new concept and at least for standalone systems some effort was already done by the research community. Here we mention some of these approaches. JXTA-SOAP project [23] was started by Kevin Burton. The JXTA-SOAP is a package which allows SOAP communication over the JXTA P2P network. Qu and Nejdl [24] discussed the exposition of existing JXTA services as web services; and also integrating web service enabled content providers into JXTA. Hajamohideen [25] discussed the use of SOAP services as web services in JXTA. The SOAP service is a customization of the peer group service with ability to send and receive SOAP messages via a JXTA transport.

Moreover from these developments the JXTA community has developed a light version of JXTA for mobile devices, called JXME (JXTA for J2ME). JXME works on MIDP supporting devices like smart phones. JXME simplified the Mobile Host's entry to P2P domain. JXME has two versions: proxyless and proxied. The proxyless version works similar to native JXTA, whereas the proxied version needs a native JXTA peer to be set up as its proxy. The proxied version is lighter of the two versions and peers using this version participate in binary communication with their proxies. Considering JXTA also eliminates many of the low level details of the P2P systems like the transportation details. The peers can communicate with each other using the best of the many network interfaces supported by the devices like Ethernet, WiFi, GPRS etc. Moreover JXTA dynamically uses either TCP or HTTP protocols to traverse network barriers, like NATs and firewalls.

Considering these advantages and features of the JXTA, the Mobile Host was adapted into the JXTA network, to check its feasibility in P2P network. Figure 1 shows the architecture of final deployment scenario of Mobile Hosts in the JXME network.

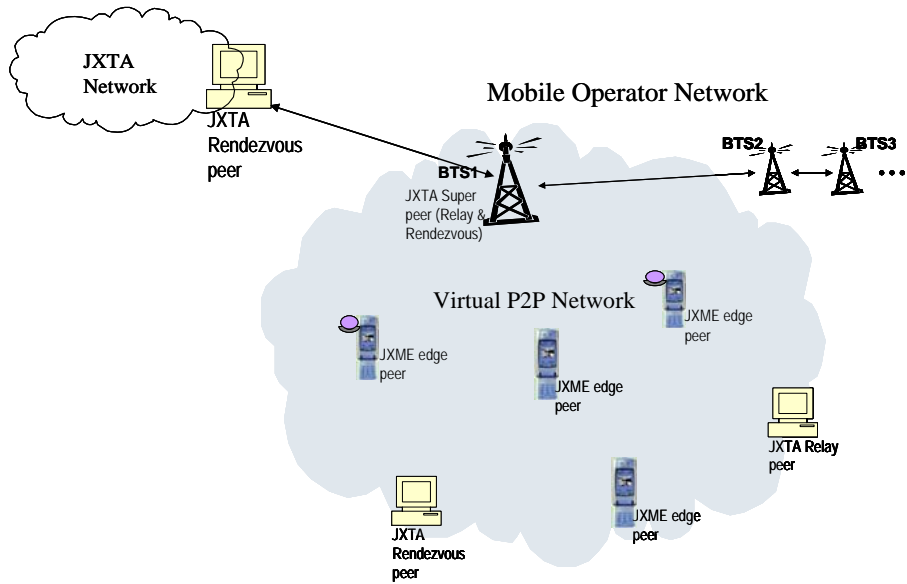

**Fig. 1.** Virtual mobile P2P network with Mobile Hosts.

As shown in figure 1, the virtual P2P network is established in the mobile operator network with one of the node in operator proprietary network, acting as a JXTA super peer. JXTA network supports three types of peers to be connected to the network. The general peers are called Edge peers. An edge peer registers itself with a rendezvous peer to connect to the JXTA network. Rendezvous peers cache and maintain an index of advertisements published by their edge peers. Rendezvous peers also participate in forwarding the discovery requests across the P2P network. A relay peer maintains route information and routes messages to peers behind firewalls. A super peer has the functionality of both relay and rendezvous peers.

The super peer can exist at Base Transceiver Station (BTS) and can be connected to other base stations extending the JXTA network into the mobile operator network. Any Mobile Host or mobile web service client in the wireless network can connect to the P2P network using the node at base station as the rendezvous peer. The super peer can also relay requests to and from JXTA network, to the smart phones. With in this network, the participating smart phones can be addressed with both peer ID and the mobile phone number, thus eliminating the need for a public IP. The necessity of public IP for each smart phone is observed to be major hindrance for the commercial success of Mobile Host, even though IPV6 [26] promises such availability. Standalone systems can also participate in such a network as both rendezvous and relay peers, if the operator network allows such functionality, further extending the P2P network.

## 4 Discovery of Mobile Web Services in JXTA Network

In JXTA the decentralization is achieved with the advertisements. All resources like peers, peer groups and the services provided by peers in JXTA network are described using Advertisements. Advertisements are language-neutral metadata structures represented as XML documents. Peers discover each other, the resources available in the network and the services provided by peers and peer groups, by searching for their corresponding advertisements. Peers may cache any of the discovered advertisements locally. Every advertisement exists with a lifetime that specifies the availability of that resource. Lifetimes gives the opportunity to control out of date resources without the need for any centralized control mechanism. To extend the life time of an advertisement, the advertisements are to be republished.

Thus to achieve alternate discovery mechanism for mobile web services, the services deployed on Mobile Host in the JXTA network are to be published as JXTA advertisements, so that they can be sensed as JXTA services among other peers. JXTA specifies 'Modules' as a generic abstraction that allows peers to describe and instantiate any type of implementation of behaviour in the JXTA world. So the mobile web services are published as JXTA modules in the P2P network. The module abstraction includes a module class, module specification, and module implementation. The module class is primarily used to advertise the existence of a behaviour. Each module class contains one or more module specifications, which contain all the information necessary to access or invoke the module. The module implementation is the implementation of a given specification. There might be more than one implementation for a given specification across different platforms.

To publish the mobile web services in the JXTA network, a standard Module Class Advertisement (MCA) is published into the P2P network, declaring the availability of a set of web service definitions, in that peer group. Once new web services are developed for the Mobile Host, the WSDL descriptions of these services are incorporated into the Module Specification Advertisements (MSA), and are published into the P2P network. The MSAs are published into JXME network with an approximate life time that specifies the amount of time the Mobile Host wants to provide the service. The MSAs are cached at rendezvous peers or any other peers, with sufficient resource capabilities. Once the life time expires the MSAs are automatically deleted from the P2P network, thus avoiding the stale advertisements. If the Mobile Host wants to extend the life time of the provided service, the MSA can be republished. The MSA can be published into the network by a service developer or even by the Mobile Host. The structure of the MSA is shown in figure 2.

```
<?xml version="1.0" encoding="UTF-8"?>
<jxta:MSA>
   <MSID> . . . </MSID>
   <Name> . . . </Name>
   <Crtr> . . . </Crtr>
   <SURI> . . . </SURI>
   <Vers> . . . </Vers>
   <Desc> . . . </Desc>
   <Parm>
```

```
        <WSDL>
            <definitions …>
                <message …> . . . </message>
                <portType …> . . . </portType>
                . . .
            </definitions>
        <WSDL>
    </Parm>
    <jxta:PipeAdvertisement> . . . </jxta:PipeAdvertisement>
    <Proxy> . . . </Proxy>
    <Auth> . . . </Auth>
</jxta:MSA>
```

**Fig. 2.** Structure of module specification advertisement (MSA) advertising a web service.

The MSA contains unique identifier (MSID) that also includes the static Module Class ID, which identifies the web services module class advertisement. The other elements of MSID include name, creator, specification and description of the advertisement. The optional element Parm consists the description (WSDL) of the web service being advertised. The PipeAdvertisement consists the advertisement of the pipe which can be used to connect to the specific web service deployed on the Mobile Host. The receiving endpoint of the pipe can be addressed with a Peer ID of the respective peer. Thus if the invocation of mobile web service is across the JXTA network, using pipes, the need for public IP is eliminated. This sort of invocation is being studied and for the time being the Mobile Host is addressed with IP and once the web services are discovered the communication between the Mobile Host and mobile web service client is still SOAP over HTTP. The remaining two elements, Proxy and Auth from MSA carry the proxy module and the security (authentication) information of the wed service module.

The module specification advertisements carrying the web service descriptions can be searched by name and description parameters. The JXTA API provides a simple keyword search on the name and description elements of the modules advertised in mobile P2P network. As we are considering about huge numbers of mobile web services, these basic parameters might not be sufficient to find out the exact search results. In fact, some valuable information like context information may not be included in these basic XML tags. Moreover we would like to extend the search criteria to the WSDL level. This means that search parameters would not be restricted to module specification advertisement details. The search will also extend by looking up the WSDL tags and information. The main idea behind this approach is that people usually express their opinion by using frequently used words and the frequency of a keyword in WSDL description is also relevant. To handle this, advanced discovery of mobile web services in P2P, index searching tools are used to match the best suited services.

This detailed search mechanism might not be performed at the JXME edge peer because of the resource limitations of the smart phones. The advanced search mechanism can be shifted to a standalone distributed middleware. In this domain, we are trying to realize an Enterprise Service Bus (ESB) [27] based "Mobile Web Services Mediation Framework" (MWSMF) [28], which maintains the individual user profiles, personalization settings and context sensitive information. ESBs are the emerging infrastructure components for realizing SOA and enterprise integration. In the scenario where the Mobile Host uses the proxied version of JXME, the proxy node can be a participant in the mediation framework, handling the discovery issues.

### 4.1 Advanced Matching/Filtering of Services

As already discussed, the basic mobile web service discovery in JXTA networks, across module specification advertisements is purely based on text based keywords. Hence the search returns a large number of resulted services, returning every service that matches the keyword. Since the discovery client in this scenario is a smart phone, the result set should be quite small so that the user can scroll through the list and can select the intended services.

Subsequently to order the JXTA search resulted services according to their relevancy, Apache Lucene tool [29] is used. Lucene is an open source project hosted by Apache and provides a Java based high-performance, full-featured text search engine library. Lucene allows to add indexing and searching capabilities to user applications. Lucene can index and make searchable any data that can be converted to a textual format. Using the tool and its index mechanism the search results were ordered/filtered and the advanced matched services were returned to the discovery client.

Modules advertising the web services in JXTA can also be properly categorized using peer groups. Web services of the same category like services of same publisher, same business type can thus be published in the same peer groups. Hierarchies of peer groups can be maintained in JXTA. Categories help in identification or classification of all the web service types and help in easy discovery of web services. The peer groups thus simulate the tModel feature of the UDDI. Currently we are studying to compare the P2P discovery approach with UDDI and are trying to join the best features of UDDI into our mobile P2P discovery approach.

Once the mobile web services were discovered from the P2P network, the mobile user can scroll through the list of services and can select the best possible service. The web service invocation client is dynamically generated at the proxy using the WSDL2Java tools [4] and is downloaded and installed into the smart phone. The deployed client software can then be used to access the service from the Mobile Host. Alternatively mobile applications were developed by composing multiple services and the applications were advertised into the P2P network and were shared using P2P file sharing mechanisms.

### 4.2 Evaluation of Mobile Web Service Discovery

To evaluate the approach, a JXTA P2P network is established with smart phones connecting to a stand alone relay peer. The relay also acts as a JXME proxy for the mobile phones and thus connecting them to the JXTA network. The relay peer is connected to a stand alone PC, which acts as a rendezvous peer. The rendezvous peer can further connect to other rendezvous peers. Thus the P2P network is established and the network is extended to public JXTA network. The JXME P2P scenario is shown in figure 3.

The mobile web services developed for the Mobile Hosts were deployed on the P910i based Mobile Hosts and the services were advertised according to the approach at the rendezvous peer1. Some of the example services used for this discovery are the mobile picture service that advertised the pictures taken by the smart phone, weather service, and some other basic services providing sensory information collected by the smart phone, used in health care systems of individuals. After publishing these basic services, alternate smart phones connected to the P2P network using the relay peer, shown in figure 2, searched for the services in the P2P network. The smart phones were successful in identifying the services in the P2P network, with reasonable performance penalties. The discovery process took less than a second for most of the services. The scalability of the approach is yet to be verified once the UDDI mapping is finalized.

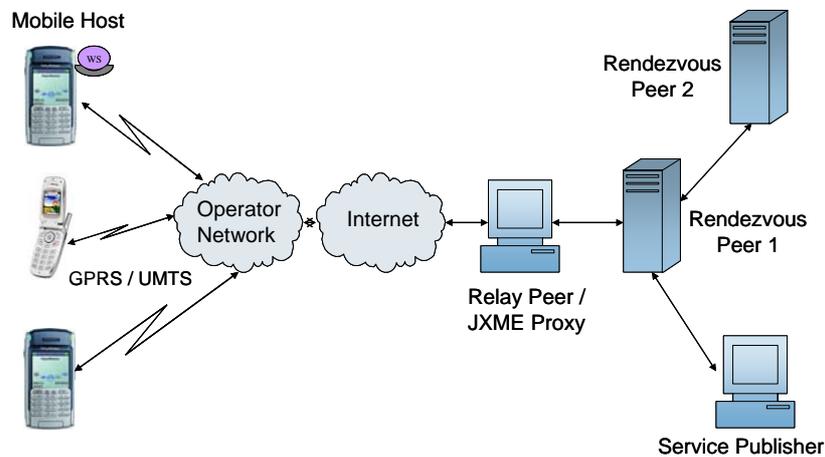

**Fig. 3.** The P2P based mobile web service discovery evaluation scenario.

Since the approach is keyword based search and the search also extended to WSDL parameters, the relevancy of the resulting services were observed to be a little indistinguishable. Mobile web service clients generally prefer using services of the Mobile Host based on several context parameters such as location, time, device capabilities, profiles, and load on the Mobile Host etc. Most of these details can not be provided just based on keywords. Once the P2P discovery approach finds its way in to the real-time environment, with each Mobile Host providing some services, providing the context information like user profiles and device capabilities is crucial in achieving much valid results.

Semantic matching of services gives the most appropriate and relevant results for mobile web service discovery. The service context and device profiles can be described using ontology-based mechanism. For describing the semantics of services, the latest research in service-oriented computing recommends the use of Web Ontology Language (OWL) [30] based Web Ontology Language for Services (OWL-S) [31]. OWL-S is an ongoing effort to enable automatic discovery, invocation, and composition of web services.

But the semantic discovery process is heavy, in terms of both resource consumption and performance latencies like extra delay. So after the analysis of our approach, we suggest using the P2P discovery mechanism first to reduce the search space. The resulted services can then be matched semantically for the most relevant results. Just as a hint, in terms of numbers, the advanced matching of services should return a set of approximately 50 services, of which the semantic matching should reduce the services to a scrollable set (5 - 10) for the smart phones.

## 5 Conclusion and Future Research Directions

The paper addresses the concept of publishing and discovery of web services deployed with Mobile Hosts in P2P networks. The approach makes use of the JXTA modules feature and provides an alternative means for discovering the mobile web services. The approach clearly solves the problem of discovering huge number of mobile web services, using resources of individual peers effectively, and at the same time eliminates the problem of inactive (stale) services.

The scalability of the approach is yet to be verified and its mapping with UDDI registry is being studied. We are also interested in extending the approach to the semantic web services domain. Basically we are looking at context aware service discovery considering device, location, and time context and user preferences of the smart phones.

Apart from the discovery mechanism, accessing the mobile web service in JXTA network, apart from the IP network is also of high interest. The access mechanism provides alternative means of addressing the Mobile Hosts and thus eliminates the need for public IP for all participating Mobile Hosts, in an operator proprietary cellular network.

**Acknowledgments.** This work is supported by German Research Foundation (DFG) as part of the Graduate School "Software for Mobile Communication Systems" at RWTH Aachen University and partly by the Research Cluster Ultra High-Speed Mobile Information and Communication (UMIC) (http://www.umic.rwth-aachen.de/)